\documentclass[12pt]{article}
\usepackage{graphicx,amssymb, amsmath}
\newcommand{\mathsym}[1]{{}}

\let\badcite=\cite
\def\cite{~\badcite}

\def\slashchar#1{\setbox0=\hbox{$#1$}           
   \dimen0=\wd0                                 
   \setbox1=\hbox{/} \dimen1=\wd1               
   \ifdim\dimen0>\dimen1                        
      \rlap{\hbox to \dimen0{\hfil/\hfil}}      
      #1                                        
   \else                                        
      \rlap{\hbox to \dimen1{\hfil$#1$\hfil}}   
      /                                         
   \fi}
    \def\slashword#1{\setbox0=\hbox{$#1$}        
  \dimen0=\wd0                                   
   \setbox1=\hbox{/} \dimen1=\wd1                
   \ifdim\dimen0>\dimen1                         
      \rlap{\hbox to \dimen0{\hfil\bf---\hfil}} %
      #1                                         %
   \else                                         
      \rlap{\hbox to \dimen1{\hfil$#1$\hfil}}    
      /                                          
    \fi}                                         %

\catcode`@=11
\newdimen\vbigd@men                             

\def\vbig#1#2{{\vbigd@men=#2\divide\vbigd@men by 2%
   \hbox{$\left#1\vbox to \vbigd@men{}\right.\n@space$}}}
\catcode`@=12

\catcode`@=11
\def\citenum#1{\csname b@#1\endcsname}
\catcode`@=12

\begin{document}
\begin{titlepage}

\bigskip\bigskip

\begin{center}{\Large\bf\boldmath
A Secure and Comparable Text Encryption Algorithm
}
\end{center}
\bigskip
\centerline{\bf Nicholas Kersting
\footnote{Email: 1054h34@gmail.com}
 }

\bigskip

\begin{abstract}

 This paper discloses a simple algorithm for encrypting text messages, based on the NP-completeness of the subset sum problem, such that the similarity between encryptions is roughly proportional to the semantic similarity between their generating messages. This allows parties to compare encrypted messages for semantic overlap without trusting an intermediary and might be applied, for example, as a means of finding scientific collaborators over the Internet.
\end{abstract}

\newpage
\pagestyle{empty}

\end{titlepage}


\section{Introduction}

There is an unavoidable trade-off between security of a text encryption scheme and its general utility:  highly-secure encryption schemes (\emph{e.g.} PGP\cite{PGP}) are only useful as a means of communication between pre-designated parties with access to the appropriate keys; at the other extreme, completely unencrypted text is optimal for search engines and public forums, but by definition obviously not secure.
One might wonder whether it is possible to find an encryption scheme that achieves a favorable compromise, being secure enough to guarantee a comfortable level of privacy while retaining some ability to sample semantic content, if only relative to other messages. This paper explicitly introduces one such scheme wherein semantic similarity between messages is deduced by direct comparison of the encrypted messages.

Communication among humans has de facto always been one of two extreme forms: exclusive privacy or public openness.
Yet in recent years there is a growing community researching a new paradigm wherein multiple parties submit private data to a collective pool which returns a public result, and this has come to be called ``Secure Multi-party Computation" (SMPC). The classic work in SMPC has to be A. Yao's ``Millionaire Problem"\cite{yao}, which defines an algorithmic procedure for two individuals to determine which among them is the richer without disclosing the individual wealth of either. In this case the private data is wealth and the public result is simply a boolean number (Yao then generalizes to m parties collectively computing a public function $f(x_1, ..., x_m)$). SMPC has most conspicuously been applied in a Danish sugar beet double auction\cite{auction} among farmers and a single buyer, wherein the private data (farmers' bids and buyer's offer) were encrypted and processed to determine an equilibrium price by matching respective supply and demand curves. Though these examples of SMPC would suggest its limitation to straight numerical applications, the premises of SMPC --- multiple parties submitting secure data which is publicly digestible to produce a particular result --- are quite general and should apply to a wide range of problems. In particular, SMPC should be a good framework for a text encryption scheme that enjoys both security and comparability. The private data in such a case would be the individual messages, while the public result might be a partitioning of all parties into groups of similar interests.

A good application of this scheme would be the problem of finding scientific collaborators. Traditionally, one attempts to find potential collaborators by proposal to selected parties, but this suffers from the fact that the more esoteric the idea, the wider the net of disclosure that must be cast to find a suitable number of potential collaborators, yet the breadth of the net is also proportional to the chance of seeding competing teams and thus potentially being `scooped'. On the other hand, if the research proposal were securely submitted in a SMPC protocol as discussed above and the public partitioning of groups indicated several parties in the group, only those parties would know something about the nature of the proposal and the chance of seeding competitors would be negligible\footnote{There is also something novel that is unachievable by traditional means here: if there were no other parties in the group, then the proposal could be certified as unique without exposing it at all.}. Members of a group could then iterate the procedure among themselves to resolve the degree of commonality, pursuing collaboration if desired.

In order to apply SMPC to the processing of text messages, however, one needs a quantitative framework that converts words into something that can be `computed'. The approach which we will employ in this paper is to define the relationships between words in a semantic network wherein words with similar meanings are neighboring nodes connected by a link, and the semantic similarity of any two words can be taken to be proportional to the minimal number of links separating them. Such a network is straightforward to construct from Princeton University's WordNet\cite{wordnet}, a publicly-available lexical database of English nouns, verbs, adverbs, and adjectives, where semantic relationships between all of these are defined systematically in a collection of cross-referenced files. In the context of the technique considered in this paper, it is not even important what the precise definitions of the words are --- the semantic network already contains enough information to quantitatively compare two messages in terms of a sort of aggregate nodal separation.

Though we shall choose English as ``the language" in this paper, it should be understood that the discussion applies to any  language in which messages are composed of discrete elements, be they words, kanji, or hieroglyphs. As tools comparable to WordNet become available in such languages, the technique of this paper will be equally applicable there.

In the next section we will describe the setup and algorithm, which for some researchers will already suffice to derive the rest of the results in this paper. Further practical results relevant to actual usage appear in Section \ref{sec:usage}, and we will go on to discuss the issue of security in Section \ref{sec:security}. Section \ref{sec:experiments} contains numerical results applied to actual messages generated with WordNet, giving a flavor of the likely usefulness of the proposed algorithm in a public setting, and we will finally consider a few generalizations and extensions in Section \ref{sec:conclusions}.

\section{Encryption}
\label{sec:encryption}
\subsection{The Dictionary and Thesaurus}
\label{subsec:dictionary}
As a preliminary step, we will need a dictionary D that maps each word in the language to a
unique integer valued in the range $[0,I_{max}]$, for some large positive integer $I_{max}$.
Also, we will need a master thesaurus T with the following two properties:
\begin{enumerate}
\item Each entry in T maps a word $w$ from D to a set of integers $\Omega_w$, called the ``synset of w". Each integer in $\Omega_w$ corresponds to a word closely related to $w$, and the j-th member of this set is denoted $\Omega_{wj}$.
\item Given any three words x, y, and z,  if x is semantically closer to y than to z, then $|\Omega_x \cap \Omega_{y}|  > |\Omega_x \cap \Omega_{z}|$.
\end{enumerate}
Whereas the first condition above is satisfied by just about any thesaurus, the second condition enforces a mathematical relation among words which allows us to gauge closeness of meaning without even knowing the language.

\begin{figure}[!htb]
\begin{center}
\includegraphics[width=5in]{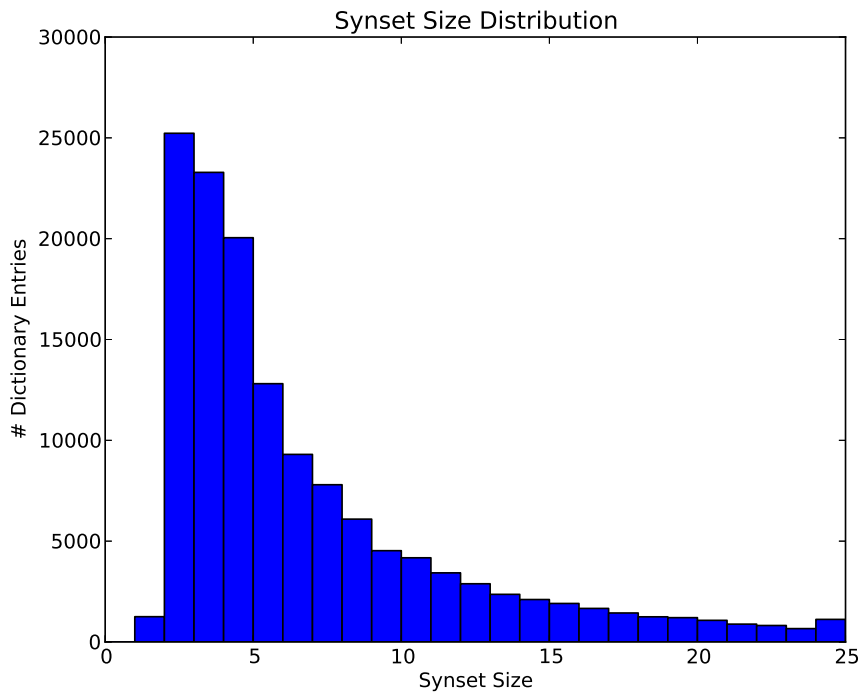}
\end{center}
\caption{\small \emph{Distribution of synsets from WordNet 3.0. Although the peak of this distribution occurs at 2, there is a very long tail (most of which is not shown) that pushes the mean synset size to 10. }
}
 \label{fig:synsets}
\end{figure}

One real example\footnote{A toy example might be a dictionary of the following three words and their synsets: x = ``dog", y = ``cat", and z = ``rock", with  $\Omega_x =  \{2, 11, 13\}$, $\Omega_y =  \{2, 11, 15\}$, and $\Omega_z = \{2, 33, 52\}$. Since ``dog" is semantically closer to ``cat" than to ``rock", $|\Omega_x \cap \Omega_{y}|  = 2 > |\Omega_x \cap \Omega_{z}| = 1$.} of T is WordNet\cite{wordnet}, representing semantic connections among roughly $1.5\cdot 10^5$ English words with $I_{max} = 2.0\cdot 10^7$. Figure \ref{fig:synsets} shows the distribution of synsets over all the words in T. Evidently the most probable synset size of a random English word in this dictionary is 2 (this includes the original word). However, as there is a very long tail\footnote{``Run", for example, has a synset of size 575, which includes integers representing ``play", ``streak", and ``function". The three largest synsets in WordNet 3.0, which in some sense represent the `most-connected' words of the English language, are ``change" (2427), ``move" (1388), and ``law" (1209).} to this distribution\footnote{A WordNet ``synset" actually comes in several different varieties, e.g. hypernym, hyponym, antonym, etc. The ``synset" defined in this paper, on the other hand, includes only words which are in direct positive relation to the parent word, and so excludes the antonym entries, for example. Also, whereas the WordNet synsets might be described as words `1-link-separated' from the parent word, the synsets in this paper consist of words `up-to-2-links-separated' from the parent word, and are thus larger.}, the mean size of a synset actually turns out to be about 10.

In addition, we will need a rule for handling words \emph{outside} D proper: although the simplest rule is to just ignore such words, a more useful possibility is to hash each such word to a unique integer not currently in the synset of any other word. If $x$ is such a word, then, $|\Omega_x| = 1$ and $|\Omega_x \cap \Omega_{y}| = 0$  for $x \ne y$.

\subsection{The Algorithm}
\label{subsec:algorithm}
Assuming that we now have a dictionary and thesaurus as defined above, we present below the  encryption algorithm in its generality, and then specialize to the simplest and most practical case for the remainder of this paper.

We will assume the ``Bag of Words" model wherein word order, grammar, and punctuation are ignored
and the message of interest is simply taken to consist of a collection  of N words with synsets $\Omega_i$  ($1 \le i \le N$). Then, for a given $n \ge 1$, the encrypted form of the message is
\begin{equation}
\label{eqn:sn}
S_n \equiv  \{\Omega_{i_{1}p_{1}} + \Omega_{i_{2}p_{2}} + ... + \Omega_{i_{n}p_{n}}\}_{1 \le i_1 < i_2 < ... < i_n \le N; ~ p_1, p_2, ... p_n}
\end{equation}
that is, the set of all integers formed by summing integers n-at-a-time, each from a unique synset\footnote{One also has the option of removing duplicate elements from $S_n$, which makes the algorithm somewhat more secure at the cost of some accuracy in utility.}. We will christen this ``n-Sum encryption".

The case of $n=2$ is the simplest case relevant to this paper\footnote{The case $n=1$ suffers from being almost trivial to decrypt, but still useful, for example, to transform a message into a list of all its words' synonyms.}, which can be rewritten as
\begin{equation}
\label{eqn:s2}
S_2 \equiv \{\Omega_{ip} + \Omega_{jq}\}_{i < j; ~ p,q}
\end{equation}
The rest of this article focuses on $S_2$, though the reader should keep in mind that the security of $S_n$ increases with $n$, and this will be investigated further in Section \ref{sec:security}.

\section{Usage}
\label{sec:usage}
\subsection{General Usage}
Usage of the n-Sum encryption algorithm would ideally follow a model wherein the only public component is a large fileserver configured such that all users have download/upload (but not overwrite) privileges. In the simplest scheme, each file, uncompressed, contains a list of integers corresponding to $S_2$ for some messsage\footnote{The filepath + name could optionally point to the message author as well.}.  Message encryption and analysis of any number of files downloaded would then be entirely client-side\footnote{In this model the server should not be trusted with these tasks, but the author makes available an explicit example combining server-side encryption, storage, and analysis\cite{qr}.}.

Once a user downloads a set of messages to analyze, there are two simple analytical operations to glean information about message content:
\begin{enumerate}
  \item Total Matching. As a first-order estimate of message similarity, it suffices to compare the intersection of the messages' sets to the sets proper:
      \begin{equation}
      \label{totalmatch}
      \xi_{S'S} \equiv \frac{|S_{2} \cap S'_2 |}{|S_{2}|}
       \end{equation}
       indicates the fractional extent to which $S$ is contained in $S'$.
  \item Word-Pair Matching. As any pair of words $(x,y)$ can define a set $S_{xy}  \equiv \{\Omega_{xp} + \Omega_{yq}\}_{p,q}$ containing $|S_{xy}| = |\Omega_x| |\Omega_{y}|$ integers, one can define a figure of merit $\zeta_{xy}$ representing the fractional semantic coverage of this word pair in the target message. Specifically,
       \begin{equation}
       \label{wordmatch}
       \zeta_{xy} \equiv \frac{|S_{xy} \cap S_2 |}{|S_{xy}|}
       \end{equation}
        Clearly if the target message contains the words x and y, then $\zeta_{xy} = 1$, while $\zeta_{xy} < 1$ indicates only partial semantic coverage of the word pair. This will be qualified by coincidental matching due to collisions with other words, the likelihood of which increases as \emph{saturation} is approached (see below).
  \end{enumerate}
  It is important to note that the only messages which will reveal any information about their content are those which already share some similarity (in the sense of (\ref{totalmatch}) or (\ref{wordmatch})) with the user's message(s).

  \subsection{Saturation}
  As each word maps to a set of integers $\Omega_w$ valued in the range $[0,I_{max}]$, $S_2$ will clearly be valued in the interval $[0,2 I_{max}]$. The above matching algorithm will start to lose effectiveness if $S_2$ densely populates this interval, for then the probability of coincidental matching approaches unity. As a quick estimate of how large a message can become before giving rise to this ``saturation" effect, note that an N-word message will contain $\frac{N (N-1)}{2}$ word pairs, and if each word has a synset of average size $\overline{\omega}$, each word pair will on average contribute $\overline{\omega}^2$ integers to $S_2$. The condition for avoiding saturation is thus
  \begin{eqnarray}{}
  \label{eqn:sat}
  \frac{N (N-1)}{2}\overline{\omega}^2 \ll 2 I_{max} \\
  \Rightarrow N \ll 2 \frac{\sqrt{I_{max}}}{\overline{\omega}}
  \end{eqnarray}
   in the large N limit ($N \gg 1$). In our WordNet setup, $\overline{\omega} \approx 10$ (see Section \ref{subsec:dictionary} above), $I_{max} = 2.0\cdot 10^7$, and thus we require $N \ll 894$. Typical ``tweet"-like messages are often closer to $N \approx 20$, and even scientific abstracts don't go much beyond $N \approx 100$ or so, thus saturation should not be a significant constraint for normal usage\footnote{One could also simply redefine D with a larger $I_{max}$ to reduce saturation effects.}. In Section \ref{sec:experiments} below, we will quantitatively verify this.

\section{Security}
\label{sec:security}
As noted previously, n-Sum encryption sacrifices some security in order to be useful as a message-comparing protocol. However, this does not mean that decryption is easy or even within reach of existing computational power; indeed it is straightforward to step up the security level of the algorithm to surpass any given computational power by increasing $n$.

Decryption of $S_n$ is almost certainly as hard as the classic NP-complete ``subset sum problem": given a set of integers X, find a non-empty subset whose sum is zero. In the present case, the elements of $S_n$ are sums of $n$ integers from D, and inversion would thus require solving a subset sum problem on D. Using all the information in $S_n$ as a series of coupled equations for an assumed number of generating words and synset sizes is probably harder than simply conducting a brute-force search on various n-word combinations from D. To crack an $S_2$ encryption, for example, one would need to compute $S_{xy}$ for each pair of words $(x,y)$ in D and compare to the encryption set $S_2$ to test for set inclusion. For the WordNet dictionary with $N = 1.5\cdot 10^5$ words, there are $\frac{N(N-1)}{2} \approx 10^{10}$ word pairs; matching against a modest-sized encryption (say a list of $10^{5}$ integers) on a typical personal computer(Intel(R) Core(TM) i7-2640M CPU @ 2.80GHz, 4.00 GB RAM) proceeds at $O(10^2)$ word pairs per second, which would thus take months of computation to find the first matching pair. Note, however, that finding subsequent pairs is much faster since one can use one of the words found in the first pair and thus need only search over $O(N)$ possible partner words in D. There are certainly ways to enhance the efficiency of this procedure, most obviously by parallel computation, but also by cataloguing all word-pairs in a database with optimized search algorithms, so $S_2$ is not expected to be highly secure in this regard\footnote{On the other hand, there are a number of things one can do to enhance security: using specialized words outside the dictionary, using a customized dictionary, or, as noted previously, eliminating duplicate entries from $S_2$ all make it harder to test set inclusion.}.

The difficulty of decrypting $S_n$ rapidly increases with $n$, however. At the next level encrypting triplets of words, a brute force attack on $S_3$ requires a search on a space already 5 orders of magnitude larger, which pushes decryption to the realm of supercomputers. $S_4$ is another 5 orders of magnitude larger and probably secure against any civilian computational power\footnote{A brute-force decryption would have to search all 4-word combinations in the dictionary, of which there are $\left( \begin{array}{c}10^5 \\ 4    \end{array} \right) \approx 10^{18}$.}. Of course, using $S_n$-level encryption requires that one has at least $n$ words in the generating message, and saturation effects also increase with $n$. Generalizing the previous result (\ref{eqn:sat}) for $S_n$, again in the large N limit, we would require
 \begin{equation}
 N \ll \frac{(2 n I_{max})^{1/n}}{\overline{\omega}}
 \end{equation}
 Of course we must also have $N \ge n$, else there wouldn't be enough words in the message to form a combination of $n$ words. For the WordNet dictionary we are using, $n = 4$ thus requires $N \le 10$, and $n > 4$ is ruled out. It turns out we can escape these constraints by simply reassigning dictionary values to increase $I_{max}$, but it is anyways useful to be aware of this in advance.

\section{Numerical Tests}
\label{sec:experiments}
A few numerical `experiments' with $S_2$ should lead credence to the some of the above claims and demonstrate practicality of n-Sum encryption as a message-comparing protocol.

\begin{figure}[!htb]
\begin{center}
\includegraphics[width=4.5in]{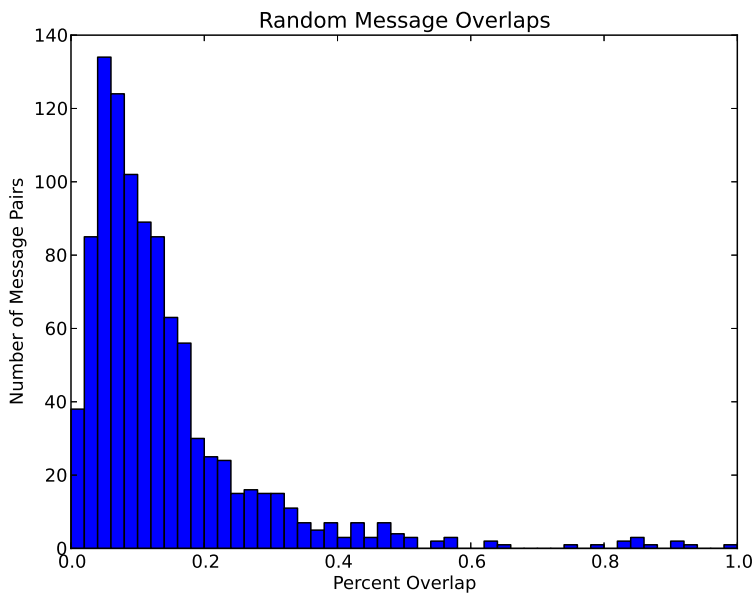}
\end{center}
\caption{\small \emph{Percent overlap ($100\times \xi_{S'S}$) between members of 1000 random pairs of uncorrelated 20-word messages. As expected, the average overlap is quite small.}
}
 \label{fig:random}
\end{figure}

First, we test to what extent random messages overlap in the Total Matching sense of Section \ref{sec:usage}. From a random sample of 1000 pairs of 20-word messages, Figure \ref{fig:random} shows that the overwhelming majority of message pairs have less than 0.5 percent overlap as defined by $\xi_{S'S}$ in (\ref{totalmatch}), with the most probable overlap being about 0.05 percent. This generally holds true for longer messages too (up to 50 words long), and only slowly approaches an average 1 percent overlap with 100-word messages. Thus, in actual usage a level of 1 percent overlap or less generally indicates unrelated messages, which makes intuitive sense. Note this also demonstrates that saturation effects are negligible for even modest-sized messages.

\begin{figure}[!htb]
\begin{center}
\includegraphics[width=4.5in]{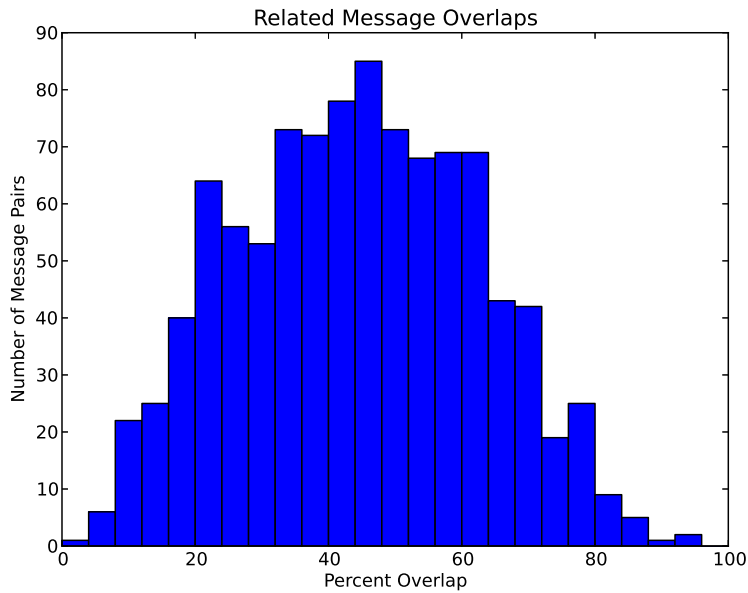}
\end{center}
\caption{\small \emph{Percent overlap between members of 1000 random pairs of related messages. The peak and average value are roughly 44 percent.}
}
 \label{fig:related}
\end{figure}

As the next check, we repeat the above analysis for a sample of 1000 pairs of \emph{related} 20-word messages, where each word in the second message in each pair is randomly chosen from one of the synsets of the first message's words. Referring to Figure \ref{fig:related}, we see that the typical message pair exhibits a significant overlap --- about 44 percent on average --- and this again makes intuitive sense. Presumably if a user were to discover overlap of this magnitude between messages $S$ and $S'$, the next step would be to identify the individual words responsible for the overlap with (\ref{wordmatch}).

Note that in these experiments the messages were randomly selected from words in D; in actual practice where a user's message may contain punctuation, special characters, and stop words, for example, one would need to parse the message into an appropriate collection of words.

Finally, let us comment on the storage space required for these messages. An N-word message encrypted with $S_n$ is essentially a list of $\frac{N^n}{n}\overline{\omega}^n$ 32-bit integers, which ought to be highly compressible, at least to the level of 1 byte per integer\footnote{For example, we can record the lowest 32-bit integer and then a 10-bit offset plus one append/separator bit per integer from there. This already gives a reduction from the naive 4 bytes per integer to about 11 bits $\approx 1.5~$bytes per integer, and there are surely more ingenious schemes one can employ.}. For $S_2$, in particular, there are thus about $50 N^2$ bytes in an N-word message, i.e. $20~kB$ for a 20-word message. Although this is clearly larger than the space required to store unencrypted text (20 words at 6 chars per word $\approx 120~$ bytes), most personal hard-drives should easily be able to store millions of messages.

\section{Conclusion}
\label{sec:conclusions}

 Recent times have witnessed greatly increased public desire for personal data privacy, yet too much encryption hinders the many services, e.g. social networking, that users expect to work quickly and effectively over the Internet.
 The method proposed in this paper attempts to strike a balance: n-Sum encryption is easily user-implementable without very sophisticated mathematics, configurably secure in a straightforward fashion, and potentially useful as a message-comparison scheme to find or expand one's social network.

 Though we mentioned how n-Sum encryption may be used in the context of scientific collaboration, there are certainly many more avenues of application: \emph{e.g.} corporate/political negotiation, community organizing, and advertising to name a few\cite{qr}. Indeed, it would not be surprising that the transition from current modes of communication based on either purely private or purely public methods to communication based upon SMPC would broadly revolutionize how humans interact with each other using the Internet\footnote{
 On an amusing note, since the message encryptions are just big lists of integers and message comparison consists of simply comparing lists for matching elements, there is a spectrographic interpretation of $S_n$: message encryptions, suitably-binned, can be interpreted as frequency spectra of, say, the visible light spectrum. Then one can think of n-Sum encryption as literally converting messages into colors in the absolute spectral sense (projection to the RGB gamut that humans perceive of as color is a further processing step\cite{qr}), where spectral similarity is positively correlated with message similarity. There may be opportunities for a hardware realization of n-Sum encryption in this regard.}.


\begin{thebibliography}{99}

\bibitem{PGP} Zimmermann, Philip (1995). ``PGP Source Code and Internals". MIT Press. ISBN 0-262-24039-4.
\bibitem{yao} Andrew Chi-Chih Yao: ``Protocols for Secure Computations (Extended Abstract)". FOCS 1982: 160-164.
\bibitem{auction} P. Bogetoft, D. Christensen, I. Damg˚ard, M. Geisler, T. Jakobsen, M. Krøigaard, J. Nielsen,
J. Nielsen, K. Nielsen, J. Pagter, M. Schwartzbach, and T. Toft. ``Secure multiparty computation
goes live".  In proceedings of: Financial Cryptography and Data Security, 13th International Conference, FC 2009, Accra Beach, Barbados.
\bibitem{wordnet} WordNet 3.0 (http://wordnet.princeton.edu/).
\bibitem{qr} Quantum Repoire (http://quantumrepoire.com). Expanded discussion of n-Sum encryption usage, plus a web server interface where one can submit messages, encrypt, and view overlap with other users' submissions.


\end{thebibliography}
\end{document}